# PYROELECTRIC AND DIELECTRIC PROPERTIES OF CALCIUM-BARIUM NIOBATE SINGLE CRYSTALS

**O.V. Malyshkina[1], V.S. Lisitsin[1], J. Dec[2], T. Łukasiewicz[3]**

[1]*Tver State University, Tver, Russia*

[2]*Institute of Materials Science, University of Silesia, Katowice, Poland*

[3]*Institute of Electronic Materials Technology, Warsaw, Poland*

Influence of the calcium concentration on the pyroelectric and dielectric properties of $Ca_xBa_{1-x}Nb_2O_6$ (CBN) crystals in a wide temperature range was studied. It is shown that calcium content ($x$) has an influence on the Curie point position but has no effect on the dielectric hysteresis loop shape and the appearance of temperature variations of dielectric permittivity, coercive field and remanent polarization. At the same time, it is found that $x$ affects the polarization profile.

Corresponding author e-mail: *Olga.Malyshkina@mail.ru*

## Introduction

Ferroelectric crystals with the tetragonal tungsten bronze structure (TBS) are widely studied objects of physics of heterogeneous media. These systems are featured in a number of useful properties including piezoelectric, electro-optic, photorefractive and pyroelectric ones [1–3]. Strontium barium niobate ($Sr_xBa_{1-x}Nb_2O_6$, SBN) crystals with the open tetragonal TBS are most widely used. During the last years large efforts are dedicated to the search of new materials with TBS on the basis of barium niobate. One of such material is calcium-barium niobate $Ca_xBa_{1-x}Nb_2O_6$ (CBN) crystals, which is better suited to potential applications because it has a much higher phase transition temperature (around 260 $^oC$) [4 – 8]. In contrast to SBN crystals, which may be grown in a wide range of $x$, the CBN materials exist in a crystalline form in a rather narrow interval of $0.2 < x < 0.4$ [4]. At the present time the main interests are focused on the study of the CBN28 composition. In spite of the discussion in the literature of its relaxor properties [6 – 8], so far there is no convincing evidence of the relaxor nature of this crystal. For example, the authors of [6] presented a temperature dependence of the dielectric permittivity having the main maximum independent of frequency, while it is only the subsidiary maximum occurring at the temperature of paraelectric phase which is frequency dependent and which is indicative of the existence of dielectric losses. At the same time, the main attribute of ferroelectric–relaxor behaviort is the frequency dependence of the dielectric maximium [2]. So it is of interest to study the CBN crystals with different calcium content keeping in view that in related SBN crystals the existence of relaxor properties depends on the concentration of Sr [2, 9, 10].

In the present work the Czochralski grown single crystals of CBN with nominal concentrations of calcium in the solution 28% (CBN28), 30% (CBN30) and 32% (CBN32) were studied. The investigated CBN single crystals were pulled from a melt along [001] crystallographic direction.

**Results and discussion**

The dielectric studies and the measuring of dielectric hysteresis loops were carried out on unpoled samples.The temperature dependence of $E_c$ and $P_r$ were studied by the Sawyer-Tower method [11]. Pyroelectric measurements were performed upon examination temperature dependence of the dielectric loops. For that the samples were poled in two ways: at room temperature or by cooling down of the paraelectric phase. It is found that the polarization profiles are independent of the method of poling. The coordinate dependence of the polarization (polarization profile) examined by the single frequency thermal square wave (TSW) method [12], which is based on the analysis of the pyroelectric signal measured under the condition of a rectangular heat flux modulation and leads to the determination of the thickness distribution of the pyroelectric coefficient. In this method, the coordinate dependence of the effective pyroelectric coefficient is determined from the time dependence of the pyroelectric response by using the following equation [12]:

$$\gamma_{eff}(x) = \frac{U(t)kT}{4R_{\mathrm{OA}}S\beta_0 W_0}\mathrm{Re}\left\{\left(\sum_{n=1}^{\infty}\frac{\sin^2(n\omega\tau/2)}{n\omega\tau/2}\frac{i}{\varphi_n^2\cdot 2t\sqrt{\alpha\pi f}}\left(1-\exp[\varphi_n(-x)]\right)\right)^{-1}\right\}, \quad (1)$$

where $\gamma_{eff}(x)$ is the effective pyroelectric coefficient of a layer with a thickness $x$, $R_{\mathrm{OA}}$ is the resistance of the operational amplifier, $T$=1/$f$ is the period, $\varphi_n = (1+i)\sqrt{n\omega/2\alpha}$, $\tau$ is the light-pulse duration, $\alpha$ is the thermal diffusion coefficient, $f$ is the heat flux modulation frequency, $d$ is the sample thickness, $S$ is the illuminated sample area, $W_0$ is the heat flux power density, $\beta_0$ is the coefficient of heat absorption, and $k$ is the thermal conductivity coefficient. Thus the basic equations of the TSW method determine the depth of temperature wave penetration at a given time *t*:

$$x(t) = \sqrt{\frac{2\alpha t}{\pi}} \quad (2)$$

and the effective pyroelectric coefficient of this layer given by Eq. 1.

The temperature dependence of the dielectric permittivity was studied at the frequencies of 100 Hz, 1, 10 and 100 kHz (Fig. 1). It may be seen at the presented dependencies that a shift of the phase transition temperature toward high-temperature region is observed with the decrease of Ca concentration in CBN crystals. This type of behavior of the dielectric maximum is, as expected, similar to that of SBN with Sr [9]. At the same time, in contrast to SBN with x > 0.5, there is no evidence for relaxor properties of the investigated CBN crystals. The temperature positions of the permittivity maxima are independent of frequency.

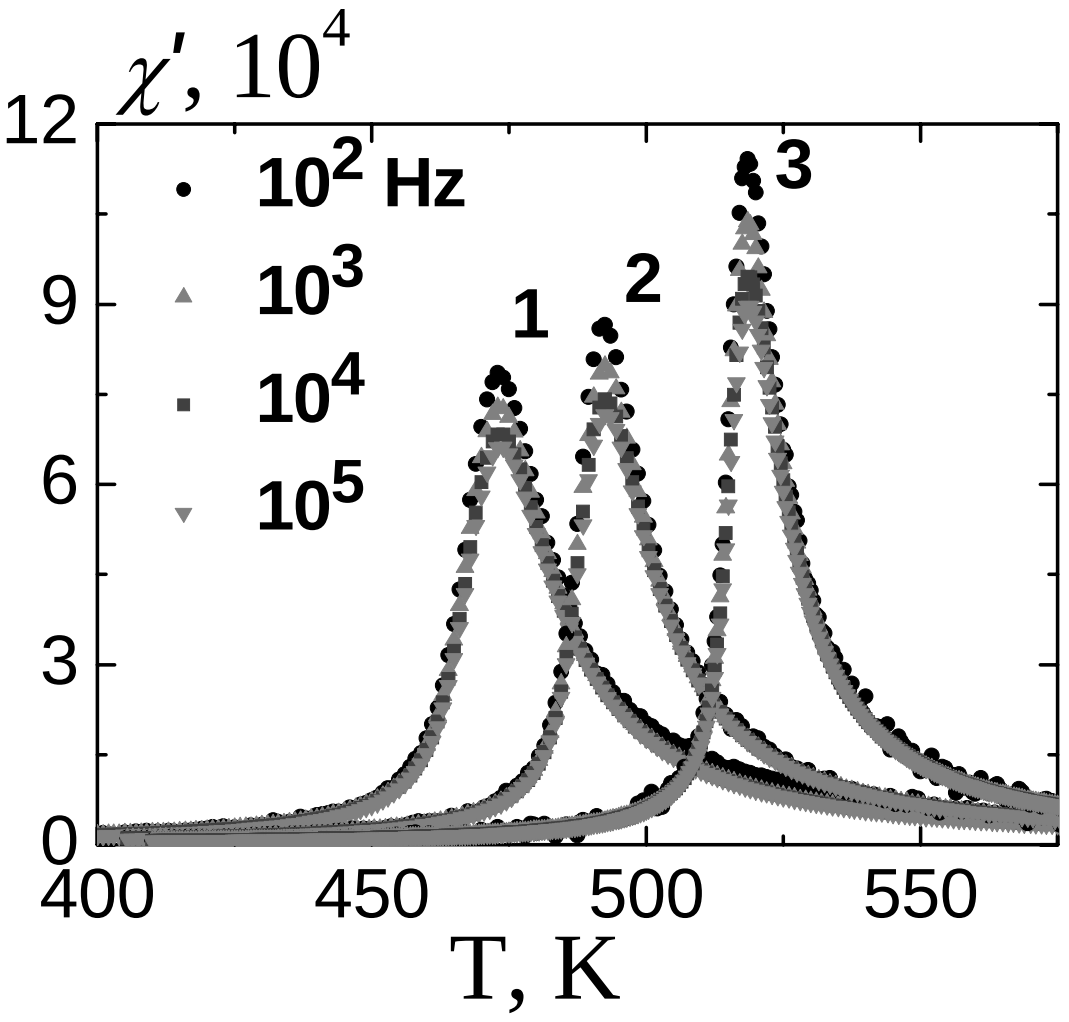

FIGURE 1. The temperature dependence of the dielectric permittivity of the CBN single crystals. Curves 1 – CBN32, 2 – CBN30, 3– CBN28

The examination of the dielectric hysteresis loop (Fig.2) has shown that the increase of the temperature results in a decrease of the coercive fields ($E_c$) accompanied by an increase of the switchable remanent polarization ($P_r$). The hysteresis loop transformation into an elliptic shape (Fig.3) was observed at temperatures of 35-50 degrees above the maximum of the dielectric permittivity (Fig.1). The observed change in shape of the loop corresponds to the sharp increase of the electric conductivity. The temperature dependence of dielectric hysteresis loop is independent of the Ca($x$) concentration.

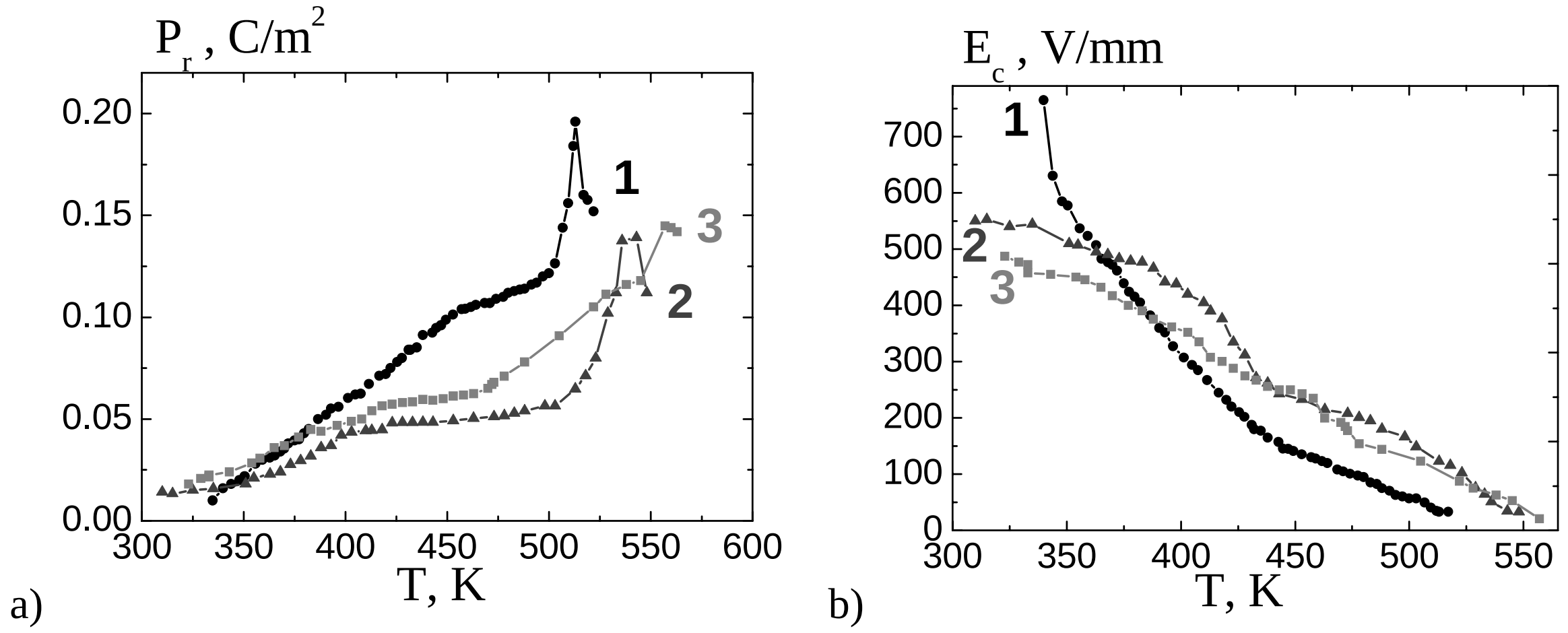

FIGURE 2. The temperature dependences of the remanent polarization (a) and the coercive field (b). Curves 1 – CBN32, 2 – CBN30, 3– CBN28

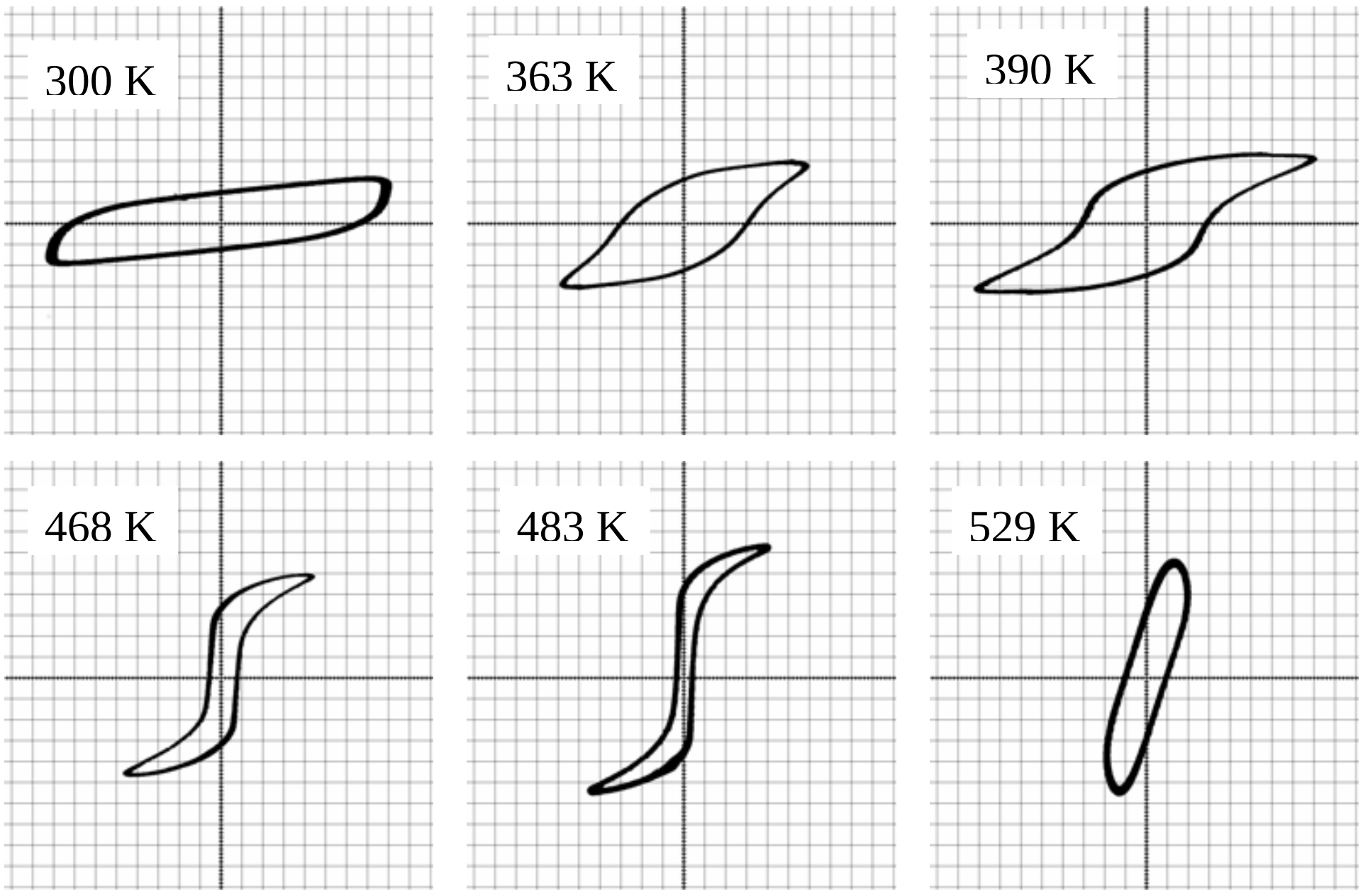

FIGURE 3. The temperature dependence of the hysteresis loop of the CBN32

In contrast to the dielectric hysteresis loop the polarization profiles (calculated using formulas (1) and (2)) of the samples depends on the Ca(*x*) concentration (Fig. 4). For all samples under studied, the polarization magnitude at the $+P_s$ side is lower than at the $–P_s$ side (Fig.4), in CBN28 the surface layer at this side is completely depolarized. It is interesting to note that the polarization profile of CBN28 reference sample not treated by AC electric field before polarization is similar to that of CBN30 (Fig. 4). After treating the CBN28 by AC field and repeated poling the polarization at the $+P_s$ side did not reappear. Thus it may be concluded that AC field results in a formation of nonuniform polarization distribution over the CBN28 crystal thickness, whereas this effect is not observed in CBN30 and CBN32.

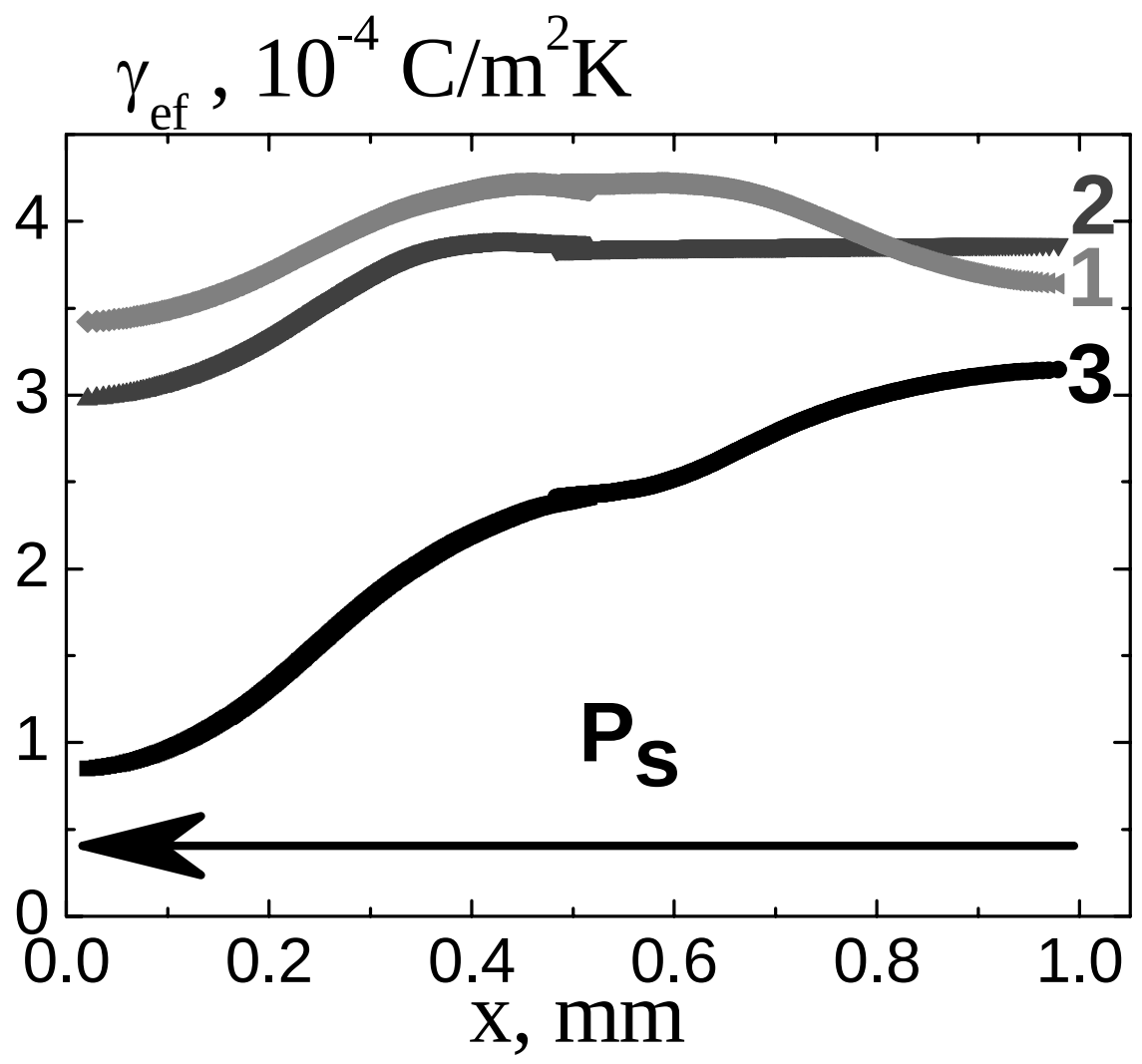

FIGURE 4 Polarization profiles of CBN crystals. Arrow shows the direction of spontaneous polarization vector. Curves 1 – CBN32, 2 – CBN30, 3– CBN28

**Conclusions**

The absence of the dielectric permittivity maximum dependence on the frequency of the probing field is indicative a lack of relaxor properties in the investigated CBN single crystals with $x = 0.28$, 030 and 0.32 (Fig. 1). At the same time the specific feature of the polarization distribution through the thickness (smaller polarization values at the $+P_s$ side) in CBN28 is similar to that of SBN61 and SBN75 crystals [10], which are known as ferroelectric relaxors. It may be supposed that the relaxor state of CBN28 discussed in [6 – 8] does not exist primordially (immediately after growth) but arises after some external influence. Since there are no intrinsic changes which may lead to nonuniform polarization distribution throughout the sample thickness, it is not unlikely that for CBN the value of $x = 0.29$ is a border line between compositions having no relaxor properties and those for which the relaxor state may be induced by external influence, in particular, by AC electric field.